\documentclass{ifacconf}

\usepackage{graphicx}      
\usepackage{natbib}        
\usepackage{xcolor}        
\usepackage{booktabs}
\usepackage{colortbl}
\usepackage{amsmath} 
\usepackage{mathtools}
\usepackage{amsfonts}
\usepackage{url}
\begin{document}
\begin{frontmatter}

\title{The \textit{E-Rocket}: Low-cost Testbed for TVC Rocket GNC Validation\thanksref{footnoteinfo}} 

\thanks[footnoteinfo]{The authors acknowledge Fundação para a Ciência e Tecnologia (FCT) for its financial support via LAETA (DOI: 10.54499/UID/50022/2025). Pedro Santos holds a Ph.D. scholarship from FCT (2023.00268.BD). © 2026 the authors. This work has been accepted to IFAC for publication under a Creative Commons Licence CC-BY-NC-ND.}

\author[First]{Pedro Santos} 
\author[Second]{André Fonte} 
\author[Second]{Pedro Martins}
\author[First]{Paulo Oliveira}

\address[First]{IDMEC and ISR, IST, University of Lisbon, 1049-001 Portugal (e-mail: \{pedrodossantos31, paulo.j.oliveira\}@tecnico.ulisboa.pt)}
\address[Second]{IST, University of Lisbon, 1049-001 Portugal (e-mail: \{andrevfonte, pedromcamartins\}@tecnico.ulisboa.pt)}

\begin{abstract}
This paper presents the \textit{E-Rocket}, an electric-powered, low-cost rocket prototype for validation of Guidance, Navigation \& Control (GNC) algorithms based on Thrust Vector Control (TVC). Relying on commercially available components and 3D printed parts, a pair of contra-rotating DC brushless motors is assembled on a servo-actuated gimbal mechanism that provides thrust vectoring capability. A custom avionics hardware and software stack is developed considering a dual computer setup which leverages the capabilities of the PX4 autopilot and the modularity of ROS 2 to accommodate for tailored GNC algorithms. The platform is validated in an indoor motion-capture arena using a baseline PID-based trajectory tracking controller. Results demonstrate accurate trajectory tracking and confirm the suitability of the \textit{E-Rocket} as a versatile testbed for rocket GNC algorithms.

\end{abstract}


\end{frontmatter}

\section{Introduction}

The space industry is undergoing a paradigm shift toward reusable launch systems for their cost-effectiveness~\citep{ragabLaunchVehicleRecovery2015} and higher launch cadence.
Innovating in this domain is recognized  as an expensive, high-risk, and elitist endeavour.
In this context, small-scale aerial vehicles emerge as accessible and safe platforms for advanced Guidance, Navigation, and Control (GNC) development and test, appealing to both industry and academia~\citep{10556959}.
Commercial off-the-shelf (COTS) prototypes, such as multirotors, have been widely adopted in GNC research for their affordability and ease of use.
However, these lack the specific dynamics and challenges of rocket flight.
The herein proposed electrically-powered rocket, \textit{E-Rocket}\footnote{All developed software and CAD files of the vehicle are publicly available at: https://github.com/The-E-Rocket-Project.}, fills this gap by providing a replicable testbed to explore novel control strategies tailored to the unique aspects of rocketry under controlled conditions.

Few scientific works address the niche of electric rockets under thrust vector control (TVC).
Notably, two works from Switzerland have been published with roots in model predictive control (MPC) and extended Kalman filter (EKF) based state estimation~\citep{spannaglDesignOptimalGuidance2021,linsenOptimalThrustVector2022a}.
Both solutions were validated in simulation and experimentally, demonstrating the agile capabilities of the platform.
Broadening the scope reveals a richer variety of strategies~\citep{Hua2013}.
Nonlinear layered architectures have been applied to sounding rockets yet lack an accessible testbed for ultimate flight validations~\citep{santosPitchPlaneTrajectory2025b}.

\begin{figure}[t]
	\centering
	\includegraphics[width=0.63\columnwidth]{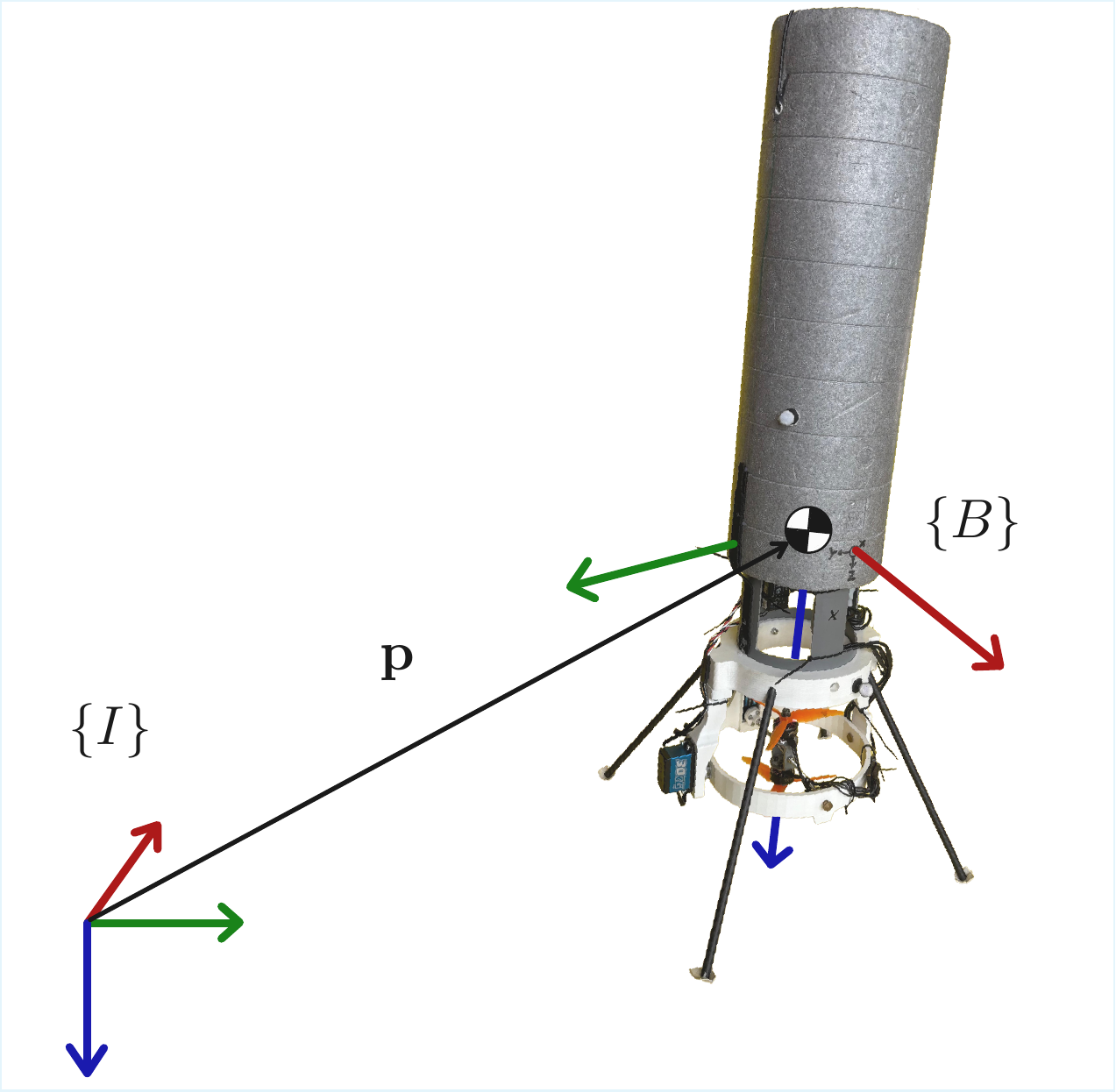} 
	\caption{The \textit{E-Rocket} (red, green, and blue correspond to the $x$, $y$, and $z$ axes, respectively).}
	\label{fig:erocket}
\end{figure}

\begin{figure}[t]
	\centering
	\includegraphics[width=0.8\columnwidth]{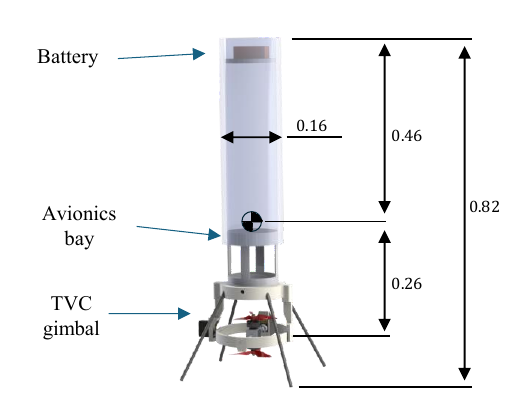} 
	\caption{\textit{E-Rocket} schematics. All dimensions in meters.}
	\label{fig:scheme}
\end{figure}

Contributions of this work span across various domains. The cost-effective mechanical design, which relies on readily available components, facilitates the  replication of the platform in academic settings. The avionics stack leverages a dual computer setup to offload low-level sensing and actuation to the well-established PX4 autopilot~\citep{paper_px4} while accommodating for custom GNC algorithms (which may have authority up to the actuators) inside a novel, open-source application running a ROS 2 network~\citep{ros2ref}. An original model is introduced, together with a PID-based control strategy for trajectory tracking, to serve as baseline for this class of vehicles. Finally, the platform is validated in an indoor flight arena with the proposed baseline control strategy.

\begin{table}[b]
	\caption{System mass properties.}
	\label{tab:parameters}
	\centering
	\renewcommand{\arraystretch}{1.1} 
	\setlength{\tabcolsep}{8pt} 
	\begin{tabular}{>{\columncolor{gray!8}}c c c}
		\rowcolor{gray!20}
		\textbf{Component} & \textbf{Value (kg)} & $\mathbf{\%}$ \textbf{Total}\\
		\hline
		 Gimbal mechanism and motors & 0.650 & 40.6\\
		 PVC tube& 0.136 & 8.5\\
	   EPS tube& 0.152 & 9.5\\
		 Carbon fiber legs& 0.033 & 2.1 \\
	   Battery & 0.294 & 18.4 \\
      Avionics & 0.338 &  21.1\\
     \hline
 Total mass & 1.6 & 100\\
		\hline
	\end{tabular}
\end{table}

\section{System Overview}
\label{sec:system_overview}

\subsection{Structural Design}
The \textit{E-Rocket} design prioritized material cost efficiency and availability, emphasizing the use of in-house 3D-printed components and readily available low-cost parts. Figure \ref{fig:scheme} presents a schematics of the \textit{E-Rocket} with some important dimensions and components identified, while Table \ref{tab:parameters} displays the system mass breakdown.

\begin{figure}[t!]
	\centering
	\includegraphics[width=0.75\columnwidth]{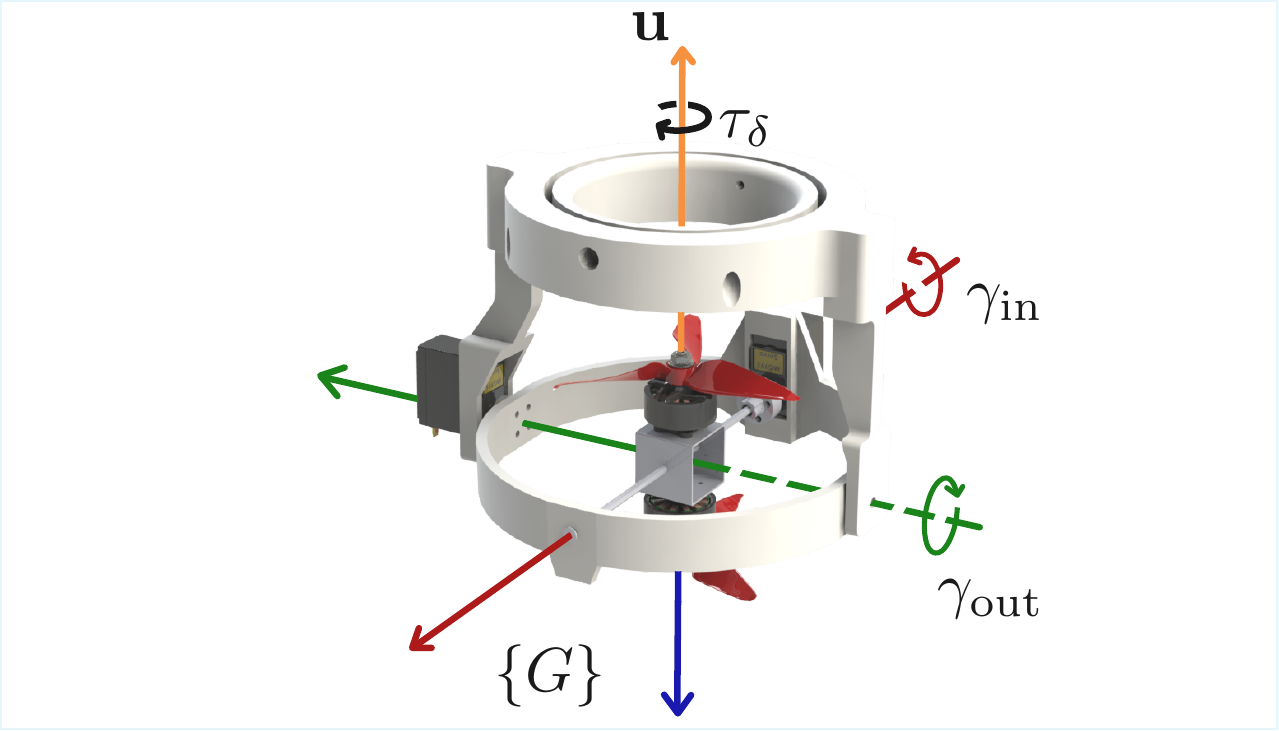} 
	\caption{Gimbal mechanism representation.}
	\label{fig:gimbal}
\end{figure}

A gimbal mechanism with two rotational degrees of freedom, represented in Fig. \ref{fig:gimbal}, is the fundamental component, which allows for Thrust Vector Control (TVC). Thrust is generated by a pair of contra-rotating brushless electric motors with 5 inch propellers. The motor pair is attached to an iron rod whose rotation is controlled by a servomotor, providing the first actuation degree of freedom. This innermost actuation mechanism is attached to a 3D printed outer ring which can rotate around an axis orthogonal to the inner one. A second servomotor controls rotation around this axis, providing the second actuation degree of freedom and completing the TVC gimbal. Each brushless motor can be controlled independently, allowing for the generation of a differential torque on the axis aligned with the thrust vector. This additional actuation input acts mainly on the yaw channel to prevent spinning motion or to actively control the yaw angle.

The gimbal mechanism is attached via two arms to a 3D printed disk that serves as core structural element of the vehicle. Four carbon fiber tubes attached to it work as legs, which allow the vehicle to take-off and land vertically on the ground. Additionally, a PVC tube, which houses the avionics bay, is mounted on top of this disk and secured in place with bolts and nuts. Ventilation ducts are present such that airflow to the motors is permitted. 

Serving as the \textit{E-Rocket}'s fuselage, the last structural component is a 0.5$\,$m long Expanded Polyestirene (EPS) tube. The battery that powers all electronic components is placed on the top part of the EPS tube. Positioning the battery at this location moves the center of mass towards the top of the vehicle, increasing the TVC lever arm and, consequently, control authority. 

\subsection{Avionics architecture}

\begin{figure*}[t!]
    \centering
    \includegraphics{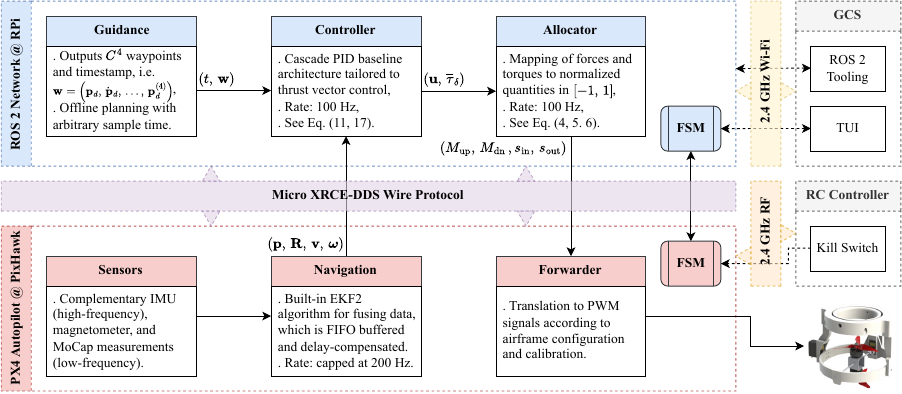}
    \caption{Proposed software partition for control and offline planning experimentation.} 
    \label{fig:software_architecture}
\end{figure*}

The avionics architecture follows a five-layer organization integrating sensing, computation, actuation, communication, and supervision:

\vspace{-1pt}
\textbf{Hardware layer -} Sensors (IMU, barometer, GNSS), actuators (servomotors and electric motors), and computing units interface with the physical system. 

\vspace{-3pt}
\textbf{Flight-control layer -} A Pixhawk flight computer runs the PX4 Autopilot to perform real-time tasks such as sensor fusion, attitude estimation, actuator control, and safety monitoring.  

\vspace{-3pt}
\textbf{Application layer -} Running on a Raspberry~Pi, ROS~2 nodes execute GNC algorithms, mission logic, and PX4 flight-mode management.  

\vspace{-3pt}
\textbf{Middleware bridge -} Two UART links interconnect the Pixhawk and Raspberry~Pi, one enabling deterministic data transfer between PX4 and ROS~2, and the other accommodating for MAVLink telemetry. 

\vspace{-3pt}
\textbf{Ground Control Station (GCS) -} The ground segment includes a Linux-based workstation running PX4 Ground Control Station (QGroundControl) and ROS~2 visualization tools. It interfaces via Wi-Fi for telemetry and mission supervision, supporting real-time monitoring, data logging, remote control through a terminal user interface (TUI), and simultaneous physics simulations.

\begin{table}[b]
    \centering
    \caption[Avionics hardware components.]{Avionics hardware components.}
    \label{tab:hardware_components}
    \renewcommand{\arraystretch}{1.1} 
	\setlength{\tabcolsep}{6pt} 
    \resizebox{\linewidth}{!}{
    \begin{tabular}{>{\columncolor{gray!8}}l c}
    \rowcolor{gray!20}
    \textbf{Component Type} & \textbf{Model}   \\
    \hline
    Flight Computer & Holybro Pixhawk 6C Mini\\
    Companion Computer & Raspberry Pi 5 Model B (8GB)\\
    Brushless Motors & T-Motor F60 PRO V LV KV2020   \\
    Propellers & Hurricane MCK 51466 V2   \\
    Electronic Speed Controllers & T-Motor AT55A    \\
    Servomotors & Power HD 30KG  \\
    RC Transmitter & FrSky Taranis X9D Plus   \\
    RC Receiver & FrSky X8R  \\
    GNSS Module & Holybro M9N GPS \\
    Power Module & Holybro PM06 V2 \\
    Lipo Battery & Tattu R-Line 1800mAh 6SP 150C\\
    \hline
    \end{tabular}
    }
\end{table}

This layered design ensures deterministic control through PX4 while retaining high-level modularity through ROS~2. The hardware configuration implements the layered architecture through COTS components, as summarized in Table~\ref{tab:hardware_components}. The electronics were selected for their proven reliability, availability, and software support.
The \textit{E-Rocket} employs a dual-computer setup, partitioned as depicted in Fig.~\ref{fig:software_architecture}, tailored to the early-stage testing of control algorithms for tracking of offline trajectories.
That said, flexibility is considered a key requirement and other configurations are trivially implemented within the network.

In our proposal, a \textit{Pixhawk 6C Mini} flight computer executes the PX4 Autopilot \citep{paper_px4} for sensor fusion, estimation, and actuator commanding, and is connected to a \textit{Holybro M9N GPS} that provides GNSS and magnetometer data. 
A \textit{Raspberry~Pi~5 Model~B} (8~GB RAM) serves as companion computer, running Ubuntu and Robot Operating System 2 (ROS~2)~\citep{ros2ref} for control, logging, and communication purposes.
Further details are provided in Section~\ref{sec:software_strategy}.
Each electric motor is driven by an Electronic Speed Controller (ESC) with an integrated 5$\,\text{V}$ Battery Eliminator Circuit. 
Paired with 5-inch propellers the motors are able to provide a combined thrust of 3$\,\text{kgf}$, sufficient for hovering at approximately 60$\%$ throttle. 
Metal-gear servomotors were selected for their torque capacity and durability. 
A power module regulates and monitors energy distribution, delivering 5$\,\text{V}$ at 3$\,\text{A}$ to avionics and measuring current and voltage for PX4 telemetry. 
The system is powered by a $6$-cell $1800\,\text{mAh}$ LiPo battery able to provide approximately three minutes of continuous hover with sufficient reserve for failsafe maneuvers. 
For safety-critical override, an RC transmitter and receiver pair provide an independent 2.4$\,\text{GHz}$ radio link connected directly to the Pixhawk S.BUS input, supporting manual intervention and arming/disarming functionality.
The architecture also supports integration with a motion-capture system that provides high-accuracy position and attitude measurements to improve state estimation accuracy. This functionality is essential for indoor testing, where {GNSS} is unavailable.

\section{Software Strategy}
\label{sec:software_strategy}

The PX4 autopilot supports most high-level UAV interactions but lacks native compatibility with thrust-vectored rockets. 
Taking advantage of the modularity of the PX4 architecture, its internal GNC pipeline was overridden by custom processes on a companion computer. 
Multiple customization levels are possible; in this work, guidance, control, and allocation were replaced with ROS 2–based components. 
PX4 thus functions primarily as a hardware abstraction layer, managing sensor acquisition, state estimation, and actuation via standard message interfaces, while the companion computer runs ROS 2 to implement GNC algorithms, mission logic, and communications.



Despite the flexibility of this modular setup, we adopt the software partition shown in Fig.~\ref{fig:software_architecture} for early-stage control experiments with custom trajectory tracking.
Reference trajectories are processed by the control cascade (Section~\ref{sec:control}) alongside state estimates from the onboard EKF. 
The control variables are translated to PWM signals through a sequence of allocation combining ROS~2 and PX4 modules, later provided to the onboard actuators.

Communication within ROS 2 relies on the DDS standard~\citep{ddsref}. A Micro XRCE-DDS agent on the companion computer bridges PX4 and ROS 2 by exposing selected PX4 publish–subscribe interfaces as ROS 2 topics and services, while also ensuring time synchronization and handling race conditions transparently.

The companion computer runs a customized ROS 2 network composed of three main nodes: (i) a Flight-Mode node that manages PX4 state transitions via validated service calls and maintains synchronization between systems; (ii) a Mission node that publishes reference trajectories and triggers mode changes according to mission phases; and (iii) a Controller node that generates actuation commands from state estimates and tracking references. For early testing, trajectories are planned offline, though the architecture can support real-time guidance methods.

\begin{figure}[t]
    \centering
    \includegraphics[width=\linewidth]{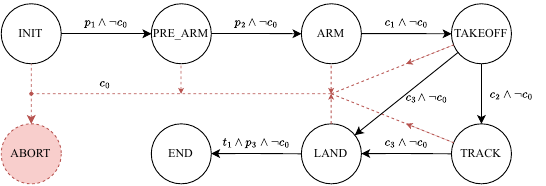}
    \caption{State machine governing flight testing phases.}
    \label{fig:state_machine}
\end{figure}

Mission profiles undergo validation prior to execution by the Flight-Mode node, which governs transitions through a finite-state machine (FSM). 
Each state represents an abstract action, with transitions triggered by validated events or internal logic.
Hierarchical fail-safes can override transitions, including heartbeat monitoring and a direct radio-controlled kill switch on PX4. 
The state workflow (Fig.~\ref{fig:state_machine}, Table~\ref{tab:logical_props}) operates alongside PX4’s internal modes, with synchronization handled via discrete events. During flight, PX4 remains in offboard mode to bypass its native controllers.
Every experimental workflow begins with a nominal deployment sequence.
After the PX4 firmware initializes in manual mode, its integration within the ROS~2 environment and external communication links must be verified.
Once both computing units are successfully armed, a takeoff command can be issued.
The prototype is designed to perform a smooth ascent to a user-defined altitude, initiating the tracking mission once commanded to.
The system is to execute a landing sequence upon mission completion, after which the state machine transitions to its ending state and requests PX4 disarming.

\begin{table}[!t]
\centering
\caption{Propositional symbols and meanings.}
\begin{tabular}{cll}
\textbf{Sym} & \textbf{Type} & \textbf{Description} \\\hline
$p_1$ & PX4 sync & Offboard request acknowledged by PX4 \\
$p_2$ & PX4 sync & Arm request acknowledged by PX4\\
$p_3$ & PX4 sync & Disarm request acknowledged by PX4 \\
$c_0$ & Command   & Abort command validated \\
$c_1$ & Command   & Takeoff command validated \\
$c_2$ & Command   & Trajectory tracking command validated \\
$c_3$ & Command   & Landing command validated \\
$t_1$ & Timeout   & Timeout landing condition reached\\\hline
\end{tabular}
\label{tab:logical_props}
\end{table}

\section{Modeling and Control}
\label{sec:control}


In order to design flight control laws for this platform, it is essential to derive a physical model that captures its dynamics and kinematics. Figures \ref{fig:erocket} and \ref{fig:gimbal} depict the three reference frames in use: an inertial frame \{$I$\}, a body frame \{$B$\} attached to the center of mass of the vehicle, and a gimbal frame \{$G$\} attached to the gimbal mechanism.   
The system has four control inputs: the inner and outer gimbal angles, $\left\{\gamma_\mathrm{in},\,\gamma_{\mathrm{out}}\right\}\in\mathbb{R}$, respectively, the combined thrust magnitude, $T\in\mathbb{R}^+$, and the magnitude of the torque caused by differential thrusting, $\tau_\delta\in\mathbb{R}$.

The gimbal angles permit expressing the thrust vector $\mathbf{u}\coloneq\left[\,u_1\:\:u_2\:\:u_3\,\right]^\mathsf{T}$ in \{$B$\} by applying two consecutive rotations, $\mathbf{R}_x(\gamma_\text{in})\rightarrow\mathbf{R}_y(\gamma_\text{out})\in\,$SO(3)  to its representation in \{$G$\}:
\begin{align}
    \mathbf{u} \coloneq &-T\,\mathbf{R}_y(\gamma_\text{out})\,\mathbf{R}_x(\gamma_\text{in})\, \mathbf{e}_3 \nonumber\\ =&-T \,\left[\,\sin\gamma_\text{out}\cos\gamma_\text{in}\:\:-\sin\gamma_\text{in}\:\:\cos\gamma_\text{out}\cos\gamma_\text{in}\,\right]^\mathsf{T}
\end{align}
where $\mathbf{e}_i$ with $i\in\{1,2,3\}$ denotes the standard orthonormal basis of $\mathbb{R}^3$. The torque input, $\boldsymbol{\tau}\in\mathbb{R}^3$, results from the thrust vectoring and differential thrusting components and is defined as
\begin{equation}
    \boldsymbol{\tau} \coloneq L\,\mathbf{S}\left(\mathbf{e}_3\right)\,\mathbf{u} + \tau_\delta\,\frac{\mathbf{u}}{\left\lVert\mathbf{u}\right\rVert}\,,
\end{equation}
where $L$ is the distance between the gimbal point and the center of mass of the vehicle and $\mathbf{S}(\cdot):\mathbb{R}^3\mapsto\mathbb{R}^{3\times3}$ is the skew operator, used for cross products.

Let $\mathbf{p} \coloneq\left[\,x\:\:y\:\:z\,\right]^\mathsf{T} \in\mathbb{R}^3$ and $\mathbf{v} \coloneq\left[\,u\:\:v\:\:w\,\right]^\mathsf{T} \in \mathbb{R}^3$ be the inertial position and velocity of the vehicle, respectively, $m\in\mathbb{R}^+$ its mass, $\mathbf{J} \in\mathbb{R}^{3\times3}$ its inertia matrix, $g\in\mathbb{R}^+$ the gravitational acceleration, $\mathbf{R}\in$ SO(3) the rotation matrix from \{$B$\} to \{$I$\}, and $\boldsymbol{\omega}\in\mathbb{R}^3$ the angular velocity of \{$B$\} with respect to \{$I$\} expressed in \{$B$\}. Then, using Newton-Euler equations for rigid body motion, the vehicle dynamics and kinematics can be expressed by the system
\begin{subequations}
\label{eq:flight_dynamics}
\begin{align}
    \dot{\mathbf{p}} = \mathbf{v} \\[5pt]
    \dot{\mathbf{v}} = g\,\mathbf{e}_3 + \frac{u_3}{m}\,\mathbf{R}\,\mathbf{e}_3 -\frac{1}{m\,L}\,\mathbf{R}\,\mathbf{S}\left(\mathbf{e}_3\right)\,\left(\boldsymbol{\tau} - \tau_\delta\,\frac{\mathbf{u}}{\left\lVert\mathbf{u}\right\rVert}\right) \label{eq:pos_dyn}\\[5pt]
    \dot{\mathbf{R}} = \mathbf{R}\,\mathbf{S}\left(\boldsymbol{\omega}\right) \\[5pt]
    \mathbf{J}\,\dot{\boldsymbol{\omega}} = -\mathbf{S}\left(\boldsymbol{\omega}\right)\,\mathbf{J}\,\boldsymbol{\omega} + \boldsymbol{\tau}
\end{align}
\end{subequations}
when neglecting aerodynamic effects and actuator dynamics.
The body is idealized axially symmetric, meaning diagonal tensor $\mathbf{J}$ with equal radial components, i.e. $j_{\perp}\coloneq j_{xx}=j_{yy}$.
Besides the underactuated nature, it is important to note the force-torque coupling appearing in the position dynamics in (\ref{eq:pos_dyn}), which can lead to undesired non-minimum phase behaviour~\citep{Hauser1992}.

\begin{figure}[t]
   \centering    
   \includegraphics[width=0.85\columnwidth]{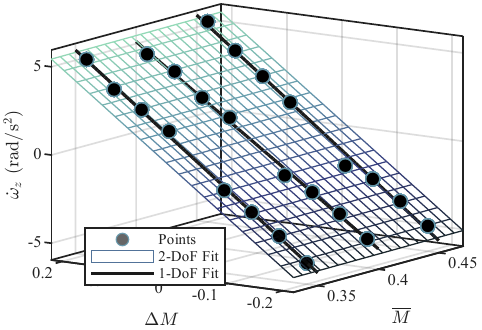}
    \caption{Linear identification of longitudinal dynamics.}
    \label{fig:torque_indentification}
\end{figure}
The system inputs must be mapped to the PX4 signals that command the actuators, requiring an identification procedure. The gimbal angles have a direct 1:1 mapping to the servo angles, which are limited to $\pm30^\circ$, leading to
\begin{equation}
    \gamma_\ell = (\pi/6)\,s_\ell\hspace{5pt}[\text{rad}]\,,\hspace{10pt}\ell\in\left\{\mathrm{in},\,\mathrm{out}\right\}\,,
\end{equation}
where $s_\ell \in [-1,\,1]$ are the PX4 servo command signals. As for the combined thrust magnitude, data from a static test lead to the linear approximation
\begin{equation}
    T = 26.45\,\overline{M} -0.3821\hspace{5pt}[\text{N}]\,,
\end{equation}
with $\overline{M} \coloneq 0.5\left(M_\mathrm{up}+M_\mathrm{dn}\right)$, where $M_\mathrm{up},\,M_\mathrm{dn} \in \left[0,\,1\right]$ are the motor command signals. Finally, the torque produced by differential thrusting was identified by measuring the angular acceleration around the z-axis for different values of the differential, $\Delta M \coloneq M_\mathrm{up}-M_\mathrm{dn}$, and average, $\overline{M}$, commands. Experimental results shown in Fig.~\ref{fig:torque_indentification} lead to the approximation on $\overline{\tau}_\delta\coloneqq\tau_\delta/j_{zz}$, written
\begin{equation}
    \overline{\tau}_\delta = 23.99\,\Delta M + 0.4031\,\overline{M} + 0.02432\hspace{5pt}[\text{rad/s}^2]\,,
    \label{eq:torque-curve}
\end{equation}
where $j_{zz}$ is the inertia value around the z-axis.
For this purpose, the obtained data was constrained to the sub-hover domain for safety.
That said, both Eq.~\eqref{eq:torque-curve}---the coefficients of $\overline{M}$ and $\Delta M$ are orders of magnitude apart---and related literature results~\citep{linsenOptimalThrustVector2022a} depict the invariance of the generated torque with the common propulsion mode.
As such, it's valid to extrapolate these results to wider domains of thrust.


To validate the proposed testbed, a preliminary control strategy for trajectory tracking was employed. 
Leveraging the principles of hierarchical and thrust direction control, a typical inner-outer solution for quadrotor UAVs is used, serving as benchmark for future, more complex algorithms. Given the small expected magnitude of the gimbal angles, two assumptions are made that allow us to consider the vehicle as fully actuated in attitude and having independent force and torque inputs: i) the force-toque coupling appearing in the position dynamics is neglected, yielding
\begin{equation}\label{eq:dyn_quad}
    \dot{\mathbf{v}} = g\,\mathbf{e}_3 + \frac{u_3}{m}\,\mathbf{R}\,\mathbf{e}_3\,,
\end{equation}
and ii) the torque input from differential thrusting is considered to act on the longitudinal axis only, i.e,
\begin{equation}
        \boldsymbol{\tau} = L\,\mathbf{S}\left(\mathbf{e}_3\right)\,\mathbf{u} + \tau_\delta\,\mathbf{e}_3\,.
\end{equation}

Consider now an inertial reference trajectory defined by $\mathbf{p}_d\left(t\right)\in\mathbb{R}^3$ of class ${C}^4$. We proceed by defining the tracking errors
\begin{equation}
\label{eq:err_sys}
    \mathbf{e}_p \coloneq \mathbf{p}_d - \mathbf{p}\,, \quad \mathbf{e}_v \coloneq \mathbf{\dot{p}}_d - \mathbf{v}\,,
\end{equation}
and a virtual acceleration input
\begin{equation}\label{eq:xi}
     \boldsymbol{\xi}\coloneq\frac{u_3}{m}\,\mathbf{R}\,\mathbf{e}_3\,,
\end{equation}
allocated through a PID control law with feedforward:
\begin{equation}\label{eq:outer_law}
   \boldsymbol{\xi}  \coloneq  \mathbf{K}_p\,\mathbf{e}_p + \mathbf{K}_d\,\mathbf{e}_v + \mathbf{K}_i\,\int\mathbf{e}_p\,dt -g\,\mathbf{e}_3 + \ddot{\mathbf{p}}_d\,,
\end{equation}
where $\left\{\mathbf{K}_p,\,\mathbf{K}_d,\,\mathbf{K}_i\right\}\in\mathbb{R}^{3\times3}$ are positive definite gain matrices. Considering the dynamics in (\ref{eq:dyn_quad}), it is a known result from the literature~\citep{Hua2013} that setting $\boldsymbol{\xi}$ as in Eq.~\eqref{eq:outer_law} renders the origin of the error system (\ref{eq:err_sys}) globally asymptotically stable (GAS) in the presence of constant disturbances. 
However, $\boldsymbol{\xi}$ is a virtual input which hides the attitude dependency in the position dynamics. 
Hence, the third component of the thrust vector is set to
\begin{equation}\label{eq:u3_law}
    u_3 \coloneq - m\,\left\lVert\boldsymbol{\xi}\right\rVert\,,
\end{equation}
and, using \eqref{eq:xi} and \eqref{eq:u3_law}, a desired attitude is defined as
\begin{equation}
    \mathbf{R}_d\,\mathbf{e}_3 \coloneq  -\frac{\boldsymbol{\xi}}{\left\lVert\boldsymbol{\xi}\right\rVert}\,.
\end{equation}
According to the ZYX Euler angle convention, with Euler angles $\boldsymbol{\lambda}\coloneq\left[\,\phi\:\:\theta\:\:\psi\,\right]^\mathsf{T}$, we have that $\mathbf{R}_d \coloneq \mathbf{R}_z(\psi_d)\mathbf{R}_y(\theta_d)\mathbf{R}_x(\phi_d)$, allowing us to write
\begin{equation}\label{eq:r3d}
    \mathbf{r}_{3d}\coloneq \mathbf{R}_y(\theta_d)\mathbf{R}_x(\phi_d)\mathbf{e}_3 = -\mathbf{R}^\mathsf{T}(\psi_d)\frac{\boldsymbol{\xi}}{\left\lVert\boldsymbol{\xi}\right\rVert}\,.
\end{equation}
From Eq. \eqref{eq:r3d}, the desired the desired roll and pitch angles can be retrieved:
\begin{equation}
    \phi_d = -\arcsin(\mathbf{e}^\mathsf{T}_2\mathbf{r}_{3d})\,,\hspace{10pt}
    \theta_d = \arctan\left(\frac{\mathbf{e}^\mathsf{T}_1\mathbf{r}_{3d}}{\mathbf{e}^\mathsf{T}_3\mathbf{r}_{3d}}\right)\,.
\end{equation}

Note that the desired yaw angle, $\psi_d$, can be set independently. A control law is now needed to ensure attitude tracking. As a preliminary approach, a single-loop PID controller is used for each angular degree of freedom. By defining a compact attitude tracking error as
\begin{equation}
    \mathbf{e}_{\boldsymbol{\lambda}} \coloneq \boldsymbol{\lambda}_d - \boldsymbol{\lambda}\,,
\end{equation}
the normalized torque input is set to
\begin{equation}\label{eq:tau_law}
    \overline{\boldsymbol{\tau}}\coloneqq\mathbf{J}^{-1}\,\boldsymbol{\tau} = \mathbf{K}_{p,\,\lambda}\,\mathbf{e}_{\boldsymbol{\lambda}} - \mathbf{K}_{d,\,\lambda}\,\boldsymbol{\omega} + \mathbf{K}_{i,\,\lambda}\,\int\mathbf{e}_{\boldsymbol{\lambda}}\,d\tau,
\end{equation}
where $\left\{\mathbf{K}_{p,\,\lambda},\, \mathbf{K}_{d,\,\lambda},\,\mathbf{K}_{i,\,\lambda}\right\}\in\mathbb{R}^{3\times3}$ are positive definite diagonal gain matrices, and the angular derivatives were approximated as body rates, i.e. $\dot{\boldsymbol{\lambda}}\approx\boldsymbol{\omega}$, which is valid for the vertical orientation the \textit{E-Rocket} maintains throughout the proposed indoor trajectory. 
Finally, using the already defined control laws, (\ref{eq:u3_law}) and (\ref{eq:tau_law}), the thrust vector input can be computed as
\begin{equation}
     \mathbf{u} =-\frac{j_\perp}{L}\,\mathbf{S}\left(\mathbf{e}_3\right)\,\overline{\boldsymbol{\tau}} +u_3\,\mathbf{e}_3\,,
\end{equation}
thus quantifying the four degrees of control of the system, described as
\begin{subequations}
\label{eq:4-dof-inputs}
\begin{align}
    T = \left\lVert\mathbf{u}\right\rVert\,,\hspace{10pt}    \overline{\tau}_\delta = {\mathbf{e}_3}^\mathsf{T}\,\boldsymbol{\overline{\boldsymbol{\tau}}}\\[5pt]
    \gamma_\mathrm{in} = \arcsin\left(\frac{u_2}{T}\right),
    \gamma_\mathrm{out} = -\arcsin\left(\frac{u_1}{\sqrt{T^2-{u_2}^2}}\right)
\end{align}
\end{subequations}
which are due the remapping to normalized quantities depicted in Fig.~\ref{fig:software_architecture}.

\section{Flight Experiments and Results}
\label{sec:experiments}


For validation, the architecture was deployed onboard the vehicle and provided with waypoint references derived from a fixed-step discretization of a ${C}^4$-continuous polynomial trajectory. These were sampled at a constant rate into timestamped vectors $\mathbf{w} \in \mathbb{R}^{ 3\times 5}$, containing position and its first four derivatives. The control laws were implemented in C\texttt{++} within the Controller node. High-fidelity software-in-the-loop simulations, using Simulink-based nodes emulating the dynamics of~\eqref{eq:flight_dynamics}, validated the system integration. Experimental flights were then conducted indoors\footnote{A video of the flight experiment here reported is available at https://www.youtube.com/watch?v=GIlzWH4nCHA.} in a $7.8\times5.0\times2.5\,$m arena equipped with an OptiTrack motion capture system, providing position and attitude measurements with accuracies of $\pm0.15\,$mm and $\pm0.5^\circ$, respectively. The system parameters were: $m=1.6\,$kg, $j_\perp = 0.375\,$kg.m$^2$, $L=0.26\,$m, $\mathbf{K}_p=\mathrm{diag}(3,\,3,\,8.5)$, $\mathbf{K}_d=\mathrm{diag}(2.5,\,2.5,\,3)$, $\mathbf{K}_i=\mathrm{diag}(0.2,\,0.2,\,1)$, $\mathbf{K}_{p,\,\lambda}=\mathrm{diag}(5,\,5,\,5)$, $\mathbf{K}_{d,\,\lambda}=\mathrm{diag}(2,\,2,\,2)$, and $\mathbf{K}_{i,\,\lambda}=\mathrm{diag}(0,\,0,\,0.5)$.

\begin{figure}
\centering
    \includegraphics{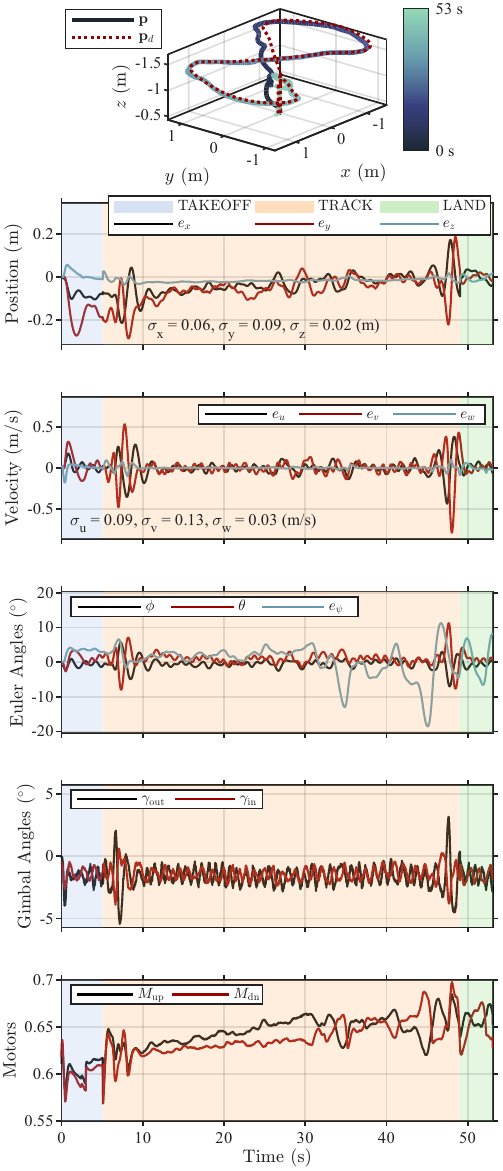}
    \caption{Time evolution of states and allocation variables.}
    \label{fig:clipped_timeframe}
\end{figure}

A 3D overview of the flight is provided in Fig.~\ref{fig:clipped_timeframe}, where the inertial position is plotted against the reference. Fig.~\ref{fig:clipped_timeframe} also illustrates the \textit{E-Rocket}'s position over time while highlighting the flight-mode transitions.
The tracking results are qualitatively satisfactory, with only small-scale errors---position-wise, the root mean squared deviations remain centimetric.
Despite the oscillations induced by the flight mode transitions, there is a clear tendency of error shrinkage.
The attitude evolution confirms the expectedly tall posture and demonstrates the propelling system's ability to contain the yaw angle near its initial value. 

Furthermore, the actuation signals evolve as expected and demonstrate the swiftness capabilities of the gimballed mechanism, despite evident room for optimizing the servo actuation.
The steady-state deviation from the origin of the gimbal angles should result from some rough calibrations and an eventual misalignment of the centre of mass with respect to the longitudinal axis.
Clear differences appear in the dynamics of $\gamma_\mathrm{in}$ and $\gamma_\mathrm{out}$ due to the asymmetric inner-outer structure, yet within the operating domains of the servomotors.
The common mode of the propellers responds more slowly, primarily focusing in counterbalancing the weight as the arena trajectory involves minimal vertical motion.
The positively-inclined evolution of the thrust command and corresponding PWM signals is a response to the voltage decay of the onboard batteries.

\section{Conclusion}
\label{sec:conclusions}

This paper presented the \textit{E-Rocket}, an electric-powered, low-cost rocket prototype for validation of GNC algorithms based on TVC. The vehicle was designed with 3D printed parts and readily available components which contribute to the cost effectiveness of the platform and yield the prototype easily replicable in academic settings. A custom avionics hardware and software stack was developed considering a dual computer setup which leverages the capabilities of the PX4 autopilot and the modularity of ROS 2. 
The platform was validated in an indoor arena using a baseline PID-based trajectory tracking controller. 
Results demonstrated accurate trajectory tracking and confirm the suitability of the \textit{E-Rocket} as a versatile testbed for rocket GNC algorithms. 
Future work includes outdoor testing, with the addition of an emergency parachute-based recovery system, and the design and validation of more advanced GNC algorithms, such as integrated nonlinear control strategies, onboard adaptive guidance, and custom navigation and state estimation methods.

\begin{ack}
The authors gratefully acknowledge the Institute for Systems and Robotics (ISR - Lisboa) for providing access to their facilities, which were essential to the research.
\end{ack}

\bibliography{ifacconf}             
                                                   







\end{document}